\documentclass[superscriptaddress,showkeys]{revtex4}

\usepackage{graphicx}
\usepackage{dcolumn}
\usepackage{bm}
\usepackage{amsmath}
\usepackage{amssymb}
\usepackage{latexsym}
\usepackage{epsfig}
\usepackage{amsbsy}
\usepackage{array}
\usepackage{subfig}
\usepackage{setspace}
\usepackage{bm}
\usepackage{textcomp}
\usepackage{epstopdf}
\usepackage{fontenc}

\begin{document}

\title{Striking Dispersion of Recrystallized Poly(ethylene glycol)-Poly(lactic acid) Solvent-Casting Blend}

\author{Xiaomin Zhu}
\author{Ran Huang \footnote{Correspondence to: ranhuang@sjtu.edu.cn}}
\author{Tian Zhong}
\affiliation{School of Chemistry and Chemical Engineering, Shanghai Jiao Tong University, Shanghai 200240, China}
\author{Ajun Wan \footnote{Correspondence to: wanajun@tongji.edu.cn}}
\affiliation{State Key Laboratory of Pollution Control and Resources Reuse, National Engineering Research Center of Facilities Agriculture, Tongji University, Shanghai 200092, China}

\date{\today}

\begin{abstract}
The solvent casted Poly(ethylene glycol) (PEG) and Poly(lactic acid) (PLA) blend was recrystallized by the thermal treatment of a heating and cooling cycle. The high PEG content blend (30\%wt) showed a striking dispersion behavior, that the materials rapidly dispersed and dissolved in the aqueous environment in a few hours. This phenomenon has not been reported by others, and was not observed in low PEG content samples of 5\%, 10\%, and 20\%, or quenched(amorphous) samples. We hypothesized the mechanism that the chain rearrangement during the thermal treatment leads to the phase separation. And with the phase separation in the recrystallized samples, the PEG potions rapidly dissolved in the aqueous environment, left out the small PLA spherulites being separated and dispersed in the solution. The same underlying reason can also be inferred from the degradation behaviors of other samples. Characterizations of DSC, XRD, and SEM have been done to validate our hypothesis.
\end{abstract}


\maketitle

\section{Introduction}
Poly(lactic acid)(PLA) has drawn intensive attentions in the last several decades as a well-recognized biodegradable, environmental friendly, and biocompatible materials \cite{rev2}. Nevertheless, comparing to the conventional petroleum-based polymers, a number of disadvantages of PLA require proper modifications for its application in either biomedical field or materials engineering \cite{rev4,rev5,rev6}. Among numerous modifiers that have been researched, the excellent biocompatibility, biodegradability, non-toxicity and chain flexibility make the Polyethylene Glycol (PEG) a good choice for the modification for PLA, in either chemical way (e.g. the PLA-PEG block polymer) \cite{pk1} or physical way (the blend mixture) \cite{Sheth,Serra,13,17,24,25,36}. Usually the PEG is used as the plasticizer to improve the mechanical properties of PLA \cite{17,36}. The phase miscibility, stability, and thermal properties of PEG-PLA materials have also been intensively investigated \cite{13,25,33,pk2}. 

On the other hand, as one of the most important properties of PLA, the degradability is an addressed concern in its modifications. Particular biological applications require either slow \cite{Albertsson1} or fast \cite{deg2} degradation especially in biomedical field. For the PEG modification on PLA, the degradation property was also investigated \cite{Sheth, Serra,24,25}, and a general conclusion is that PEG content will increase the degradation rate of PLA due to its flexibility, hydrophilicity and wettability to degrade in aqueous environment \cite{cooper}. Furthermore, although not being confirmed by any report, a significant clue can be inferred from previous researches \cite{25} that not only the PEG property itself, but also the immiscibility plays an important role in this degradation rate enhancement. Nevertheless, by this time there was an interesting gap in the research on the PLA-PEG system: almost all the studies on the PLA-PEG system involving thermal treatment, which usually leads to considerable immiscibility, were focusing on the mechanical applications and not concerning much on the degradability, while the studies on the degradation of PLA-PEG system were often regarding to the biomedical applications at a moderate temperature, e.g. $37.5^\circ C$.

Recently we developed a solvent-casted PLA tissue engineering scaffold involving a thermal treatment to adjust its crystallization \cite{RanML}, and utilized PLA-PEG blend as raw materials to adjust the degradability \cite{RanMRX}. An abnormal striking dispersion phenomenon of the recrystallized PEG-PLA solvent-casted scaffold was observed, a recrystallized sample with high PEG content of 30\%wt was found to disperse and dissolve in the aqueous environment in a few hours, while normally it takes much longer time (weeks or months) for a total degradation for either amorphous or crystallized PLA. To our best knowledge, this phenomenon has not been reported by others before. Therefore, we have conducted a set of experiments on the degradations of recrystallized PLA-PEG solvent-casted blends, and characterized the blends by XRD, SEM and DSC to figure out the underlying reason. The results firmly evidenced our hypothesis on the mechanism. Researches on the similar system reported by others \cite{Sheth, Serra} were also compared with our work, and we have confirmed the same mechanism from their results, which however were not concluded there.
\section{Experimental}

\subsection{Materials}

The PLA of label 4032D is purchased from Natureworks\circledR, with L/D
ratios from $24$:$1$ to $32$:$1$ and weight-average molecular weight $Mw\sim200,000$. The PEG 4000 ($Mw\sim4,000\pm200$) is purchased from Sinopharm\circledR (China). The analytical grade of dichloromethane is used as solvent.

\subsection{PEG-PLA Blend Preparation}

The PLA and PEG pellets were solved in dichloromethane and rigorously blended by magnetic stirrer at 600rpm for $0.5{h}$ to get a clear solution. Then the PEG-PLA solution is casted on Teflon plate and dried in vacuum for $12{h}$ to get the mixture film. The film should be observed with the uniform color and texture implying that two polymers were well-mixed. Four sets of samples labeled as A, B, C, and D with the PEG weight percentage of $5\%$, $10\%$, $20\%$, and $30\%$ are prepared for thermal treatment and characterizations. 

\subsection{Thermal Treatment}

For each wt\% sample, two blend films were heated in oven at $170{{}^{\circ}{\rm C}}$ for $0.5{h}$, then one sample was immediately quenched in liquid nitrogen to obtain the amorphous film; the other sample was recrystallized with slow cooling down at the rate of $10{{}^{\circ}{\rm C}}/30{min}$ to the room temperature. The quenched samples were labeled with apostrophe, for example the recrystallized 5\%wt sample was labeled as A while the quenched 5\%wt sample is A'.

\subsection{Characterization}

The crystallinity was measured by X-ray Diffraction (D/max-2200/PC, Rigaku Corporation). The thermal analysis was performed on DSC (Q2000, TA Instruments). The macro-structure were
revealed by Scanning Electron Microscope (SEM) (Nova NanoSEM 450, FEI).

\subsection{Dispersion/Degradation}
The dispersion/degradation of films in size of $3cm\times3cm$ were done in $ 0.1 mol/L $ NaOH solution at 37.5\textcelsius. The samples were freeze-dried for $2{h}$ and weighed every several hours.  

\section{Results and Discussion}

The characterization of XRD, DSC and SEM, and degradation experiments have been done to investigate the properties of the blends with thermal treatment. We also characterized the original blends without thermal-treatment for the control group, and found their physical and degradation properties show no obvious deviation from others' previous works. Therefore those results are not included in this paper.

\subsection{Crystallinity Feature}

\begin{figure}
    \centering{
	\includegraphics[width=0.48\textwidth]{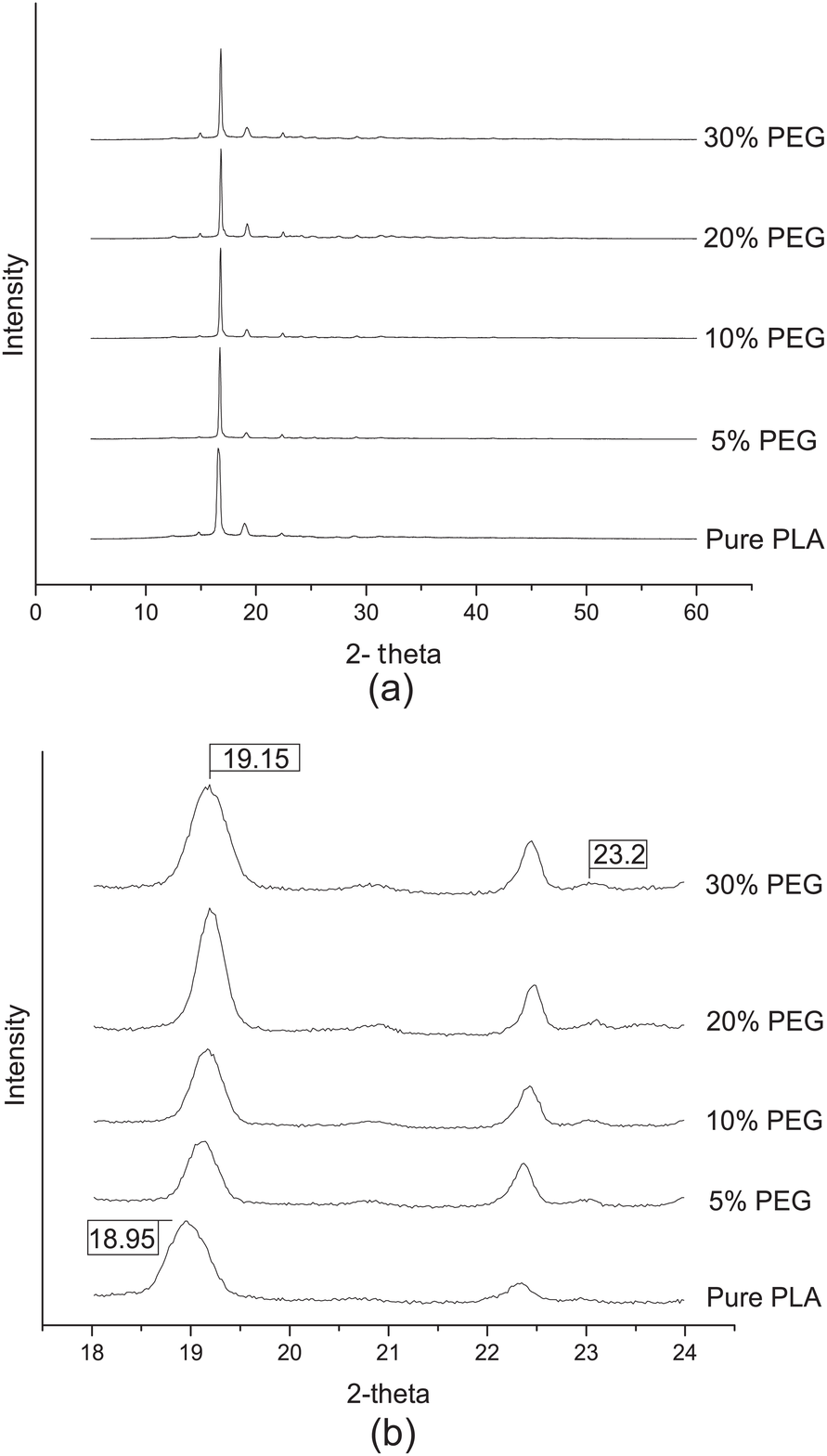}
    }
    \caption{The XRD results of Pure PLA and sample A, B, C, and D: (a) The XRD comparison of pure PLA and four blend samples. The intensity of each result is normalized with the total area; (b) The detailed window of $2\theta\in[18,24]$ to observe the effects of PEG. The pure PLA XRD data is from our previous study.}
    \label{fig:XRD}
\end{figure}

The recrystallized PEG-PLA shows similar crystallinity behavior to the Pure PLA as shown in Fig.\ref{fig:XRD}a. For comparison the XRD curves are normalized with the total area. Almost identical behaviors of blends to the pure PLA sample were obtained. Besides the main crystal peaks at $2\theta=16.5^\circ$, small differences on minor peaks are acceptable and not considered as the effects of blending. The quenched sample features total amorphous without either PLA or PEG crystal peaks thus not included here.

To observe the presentation of PEG, Fig.\ref{fig:XRD}b shows the detailed curves in the window of $2\theta\in[18^\circ,24^\circ]$. The main crystallinity peaks of PEG were reported to be at $2\theta\simeq19.1^\circ$ and $23.2^\circ$ \cite{pegXRD}. In Fig.\ref{fig:XRD}b it can be observed that, the pure PLA has a crystal peak at $\sim18.95^\circ$ and in the blends this peak is shifted to $\sim19.15^\circ$, and although tiny but visible signal at $\sim23.2^\circ$ can be observed especially with larger PEG wt\% content. Therefore from the XRD we can confirm that: 1) The crystallization of PLA is not affected by the PEG and forms semi-crystal similar to the pure sample; 2) for each wt\% content the crystallinity of PEG is very small, while its presentation can still be observed on the XRD results. For recrystallized blends, basically we have semi-crystal PLA and amorphous PEG mixtures, and that directs to the underlying reason to understand the unusual striking degradation phenomenon, which will be detailed in the following sections.  

\subsection{Thermal Properties}
\begin{figure}
    \centering{
	\includegraphics[width=0.48\textwidth]{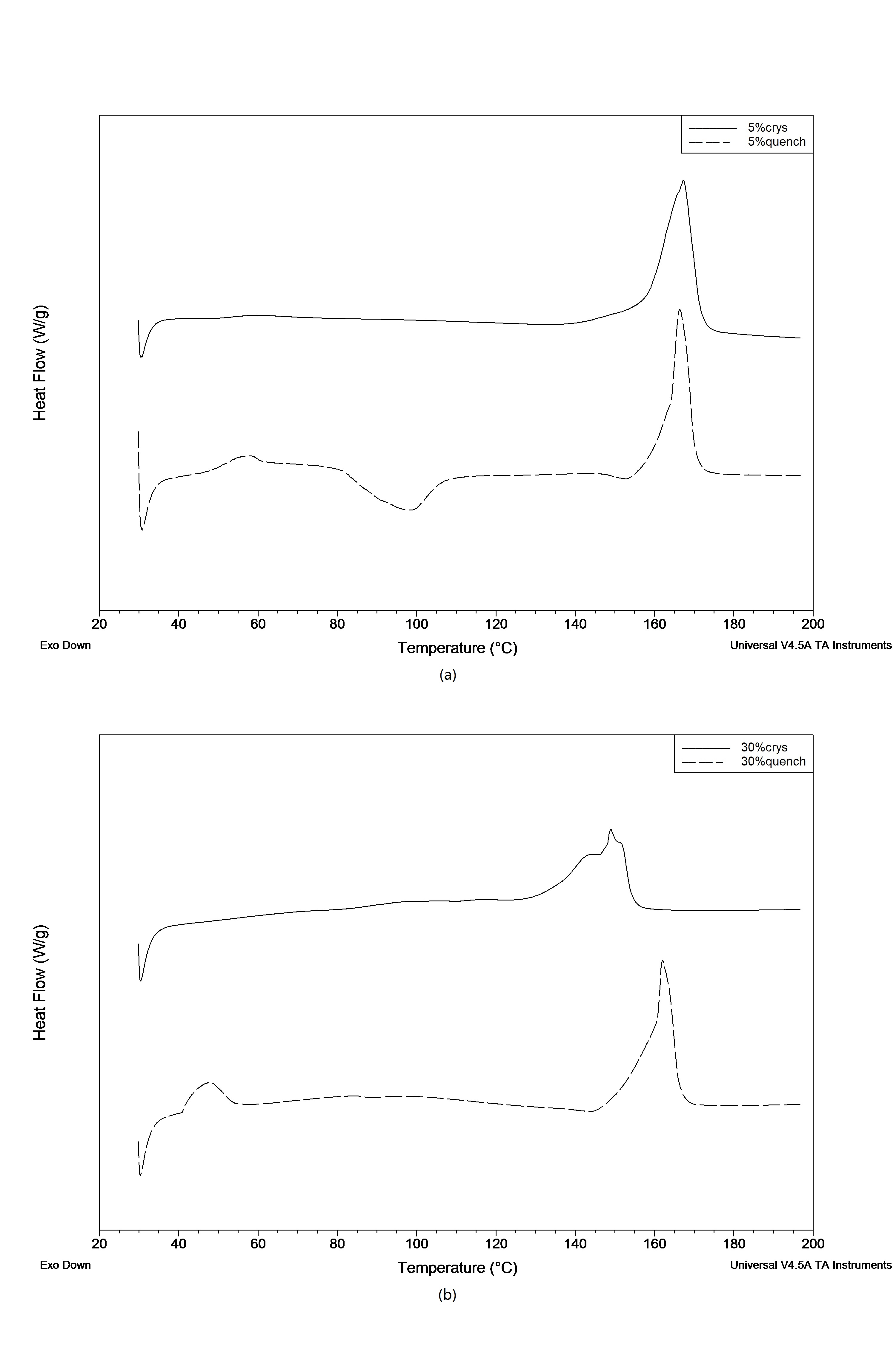}
    }
    \caption{The DSC results of sample (a) A\&A' and (b) D\&D'. All curves are the heating up process. The thermal cycle cannot be done because our samples are pre-thermal-treated. The solid line is recrystallized sample, the discontinuous line is quenched sample.}
    \label{fig:DSC}
\end{figure}

We have done the DSC on samples A\&A' and D\&D' to investigate the possible reason of striking degradation of the highest PEG wt\% materials comparing to the least PEG wt\%. Fig.\ref{fig:DSC} shows the heat flow in the heating process of sample A\&A' and D\&D'. Since our samples are thermally pre-treated, the cooling runs were not done. There are several observations in the DSC results: 1) The melting of PEG is clear in the quenched samples A' and D', while in the recrystallized samples A and D there is no such melting peak. The probable reason is that, in A and D the PEG is confined by the well-arranged PLA crystals and the heat intake can be absorbed by the crystal without phase transition. Therefore the PEG solid maintains a `super-heated' state within a long temperature region, i.e. the melting heat absorption peak is stretched to be non-observable; 2) Following the ``PLA crystal confining PEG" case, the presence of PEG significantly reduces the transition temperatures in sample D, this is consistent to other groups' reports (Ref. \cite{17} as an example). This reduction does not exist in sample A because of the low wt\% of PEG; 3) Consequently, the $T_{m}$ reduction of sample D' is small. The reason is that the PEG portion has melted at $\sim$50\textcelsius, its effect is then dismissed when the PLA is melting; 4) The glass transition temperature $T_{g}$ of PLA overlaps the melting temperature of PEG at $\sim$50\textcelsius. Thus, in the recrystallized samples A and D, the glass transition of PLA is not obvious and it covers up the melting of PEG, while in the quenched sample A' and D', the PLA is amorphous and a combined peak of PLA's glass and PEG's melting transition can be observed. It also can be seen that this peak is at 48\textcelsius in D', which is closer to the $T_{m}$ of PEG 4000 ($60^\circ$). In sample A', this peak is shifted to 57\textcelsius, which is closer to the $T_{g}$ of PLA. This shift can be explained by the correlation between two polymer chains. It seems to be counter-intuitive that the correlation is less with higher PEG wt\% as sample D' has 30\% PEG, since in normal sense there should be more interactions between two ingredients inside the matrix, nevertheless, this implies an important clue of the phase-separation and will be validated by other characterizations.

\subsection{Dispersion/Degradation Behavior}

The immersion experiments on each sample have been done several times, the degradation behavior differs however the principles are consistent. Fig.3 shows one typical degradation behavior of each sample, indicated by the weight loss percentage. Basically the degradation of samples A\&A', B\&B', C\&C', and D' are similar, that the weight loss is fast at the beginning (in the first 24 hours), and then followed by a slow and stable weight loss rate. For samples A\&A', B\&B', and C\&C', the amorphous materials always degrade faster than the recrystallized one, i.e. a sharper weight loss at the beginning and a larger slope in the following rate. Also, the higher PEG wt\% content is, the faster the sample degrades. These observations agree with others' previous research that the PEG will enhance the degradation rate and the amorphous state is easier to degrade.

\begin{widetext}

\begin{figure}
    \centering{
	\includegraphics[width=0.8\textwidth]{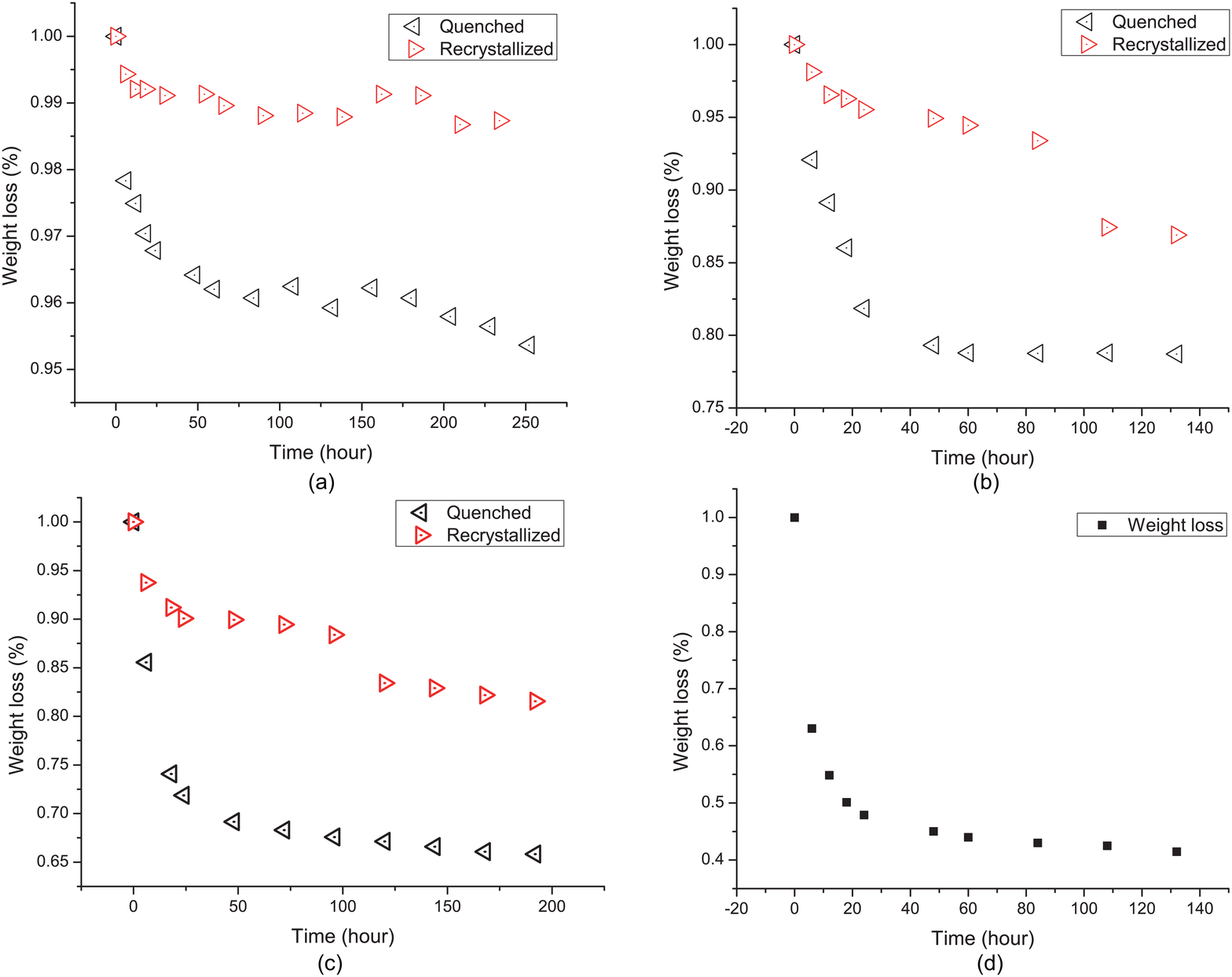}
	\caption{The degradation weight loss percentage of (a) Sample A\&A': 5\% PEG quenched and recrystallized; (b) Sample B\&B': 10\% PEG quenched and recrystallized; (c)Sample C\&C': 20\% PEG quenched and recrystallized; (d) Sample D': 30\% PEG quenched.}
    }
    \label{fig:degrad}
\end{figure}

\end{widetext}

\begin{figure}
    \centering{
	\includegraphics[width=0.48\textwidth]{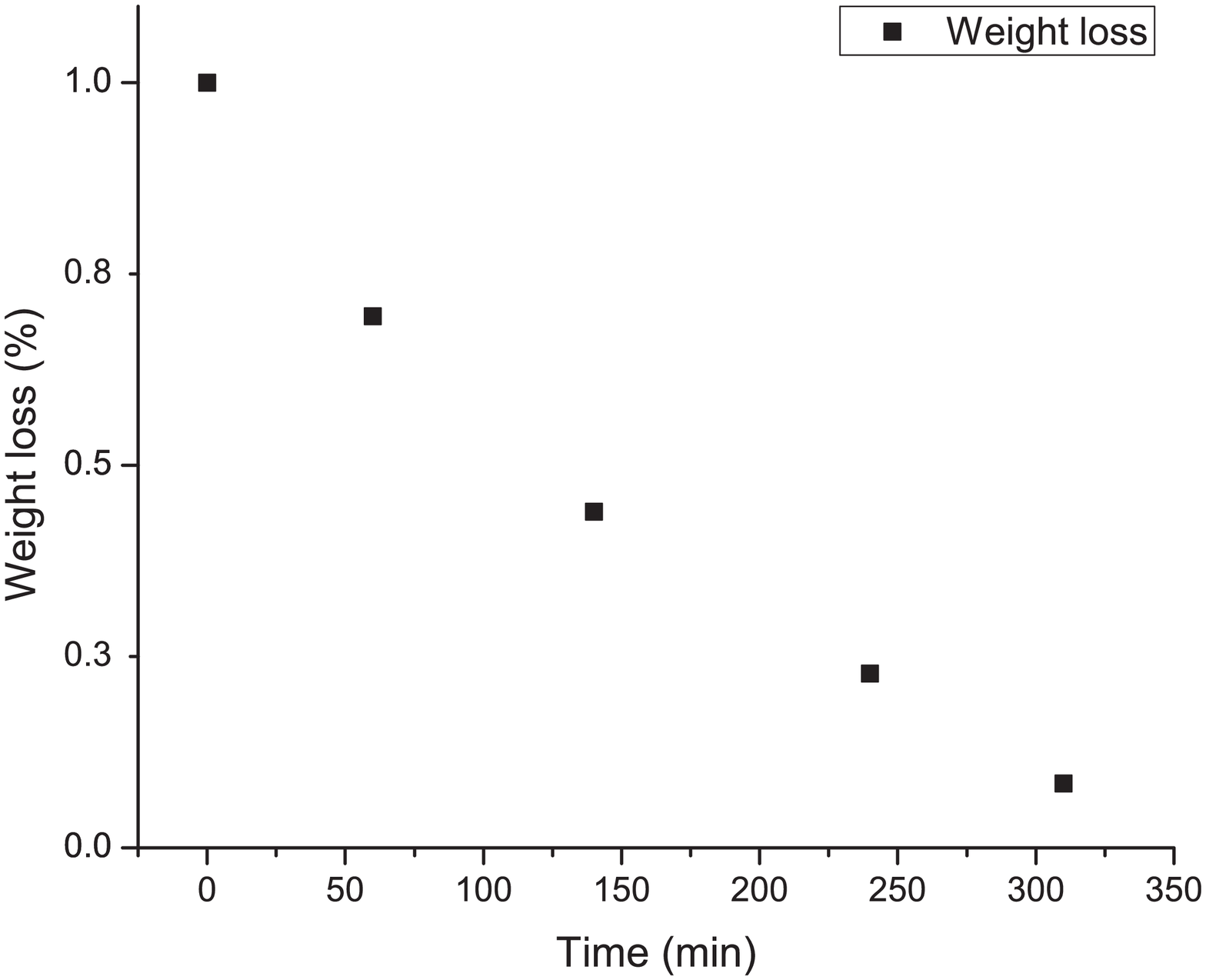}
    }
    \caption{The degradation weight loss percentage of Sample D: 30\%PEG recrystallized. The degradation was sharp and the sample dissolved in around 6 hours. }
    \label{fig:30crys}
\end{figure}

The interesting phenomenon, which is the key point of this paper, is the immersion of recrystallized 30\%wt PEG sample D. The first time we immersed the sample and tried to weight it after 8 hours, it was found that had already disappeared in the solution. Therefore, a more precious monitoring by hours has been done for the samples, several sample Ds were immersed in the solution and taken out to weight after every 1 or $2{h}$. All the samples totally dissolved in $6$-$8{h}$ and one typical result is shown in Fig.\ref{fig:30crys}.
 
We firstly found this phenomenon in the NaOH solution. The reason we select NaOH solution is for a fast degradation because in water environment it typically takes months to observe a significant weight loss. While this striking degradation of sample D was a surprise, herein we also did the immersion of sample D in water to check if it is due to the effect of alkaline environment. However similar behavior was observed, that even in water environment the sample still disappeared within $8{h}$. 

By all the means the PLA and PEG chains should not degrade in such a short time. Considering the PEG is water-solvable, a reasonable hypothesis for this phenomenon is that, for high PEG wt\% materials, the recrystallization process induces a high degree phase separation, the PLA forms spherulites or other self-aggregation crystal structure, and are separated by the PEG portions. When soaked in the aqueous environment, the PEG solves in around $6{h}$ and the small PLA portions are subsequently dispersed into the solution, these PLA portions are as small as invisible so it looks like a total degradation and the sample `disappears'. The self-aggregation of crystal is an important factor while the amorphous sample D' does not behavior so. This hypothesis agrees with the phase-separation indicated by the DSC results, i.e. PEG solving is impossible if two polymers are well mixed and entangled with each other.

We then also have explanations on some abnormal behaviors of sample A, B and C. In Fig.3a the recrystallized curve has a weight gain at $162{h}$. Remember that in the DSC analysis we made a guess on the recrystallized sample that the PEG is well-confined by PLA crystals, therefore it is possible that some PEG portions is confined and cannot solve out, but the swelling due to its wettability can cause a minor weight gain. This swelling is hard to be dehydrated by 4 hours freeze-drying. In Fig.3b and c, the recrystallized curve of B and C experience a step weight loss at 108 and $120{h}$ respectively. Because of the phase separation, the materials texture is not uniform, and partial PLA particles cracking and dispersion with the PEG solving is expectable. To check this, five samples of B and C were degraded and this step weight loss has been observed several times at some time point, Fig.3b and c are one example of this behavior.

\subsection{Morphology of Dispersion/Degradation}

We have characterized the samples B\&B', C\&C' and D\&D' by SEM to reveal the degradation morphology and to validate the hypothesis. The samples A\&A' of PEG/PLA 5/95 were not observed because their morphology should not be far away from the Pure PLA, and not helpful to investigate the striking degradation of the high PEG wt\% sample. 

\begin{figure}
    \centering{
    \subfloat[]{
	\includegraphics[width=0.48\textwidth]{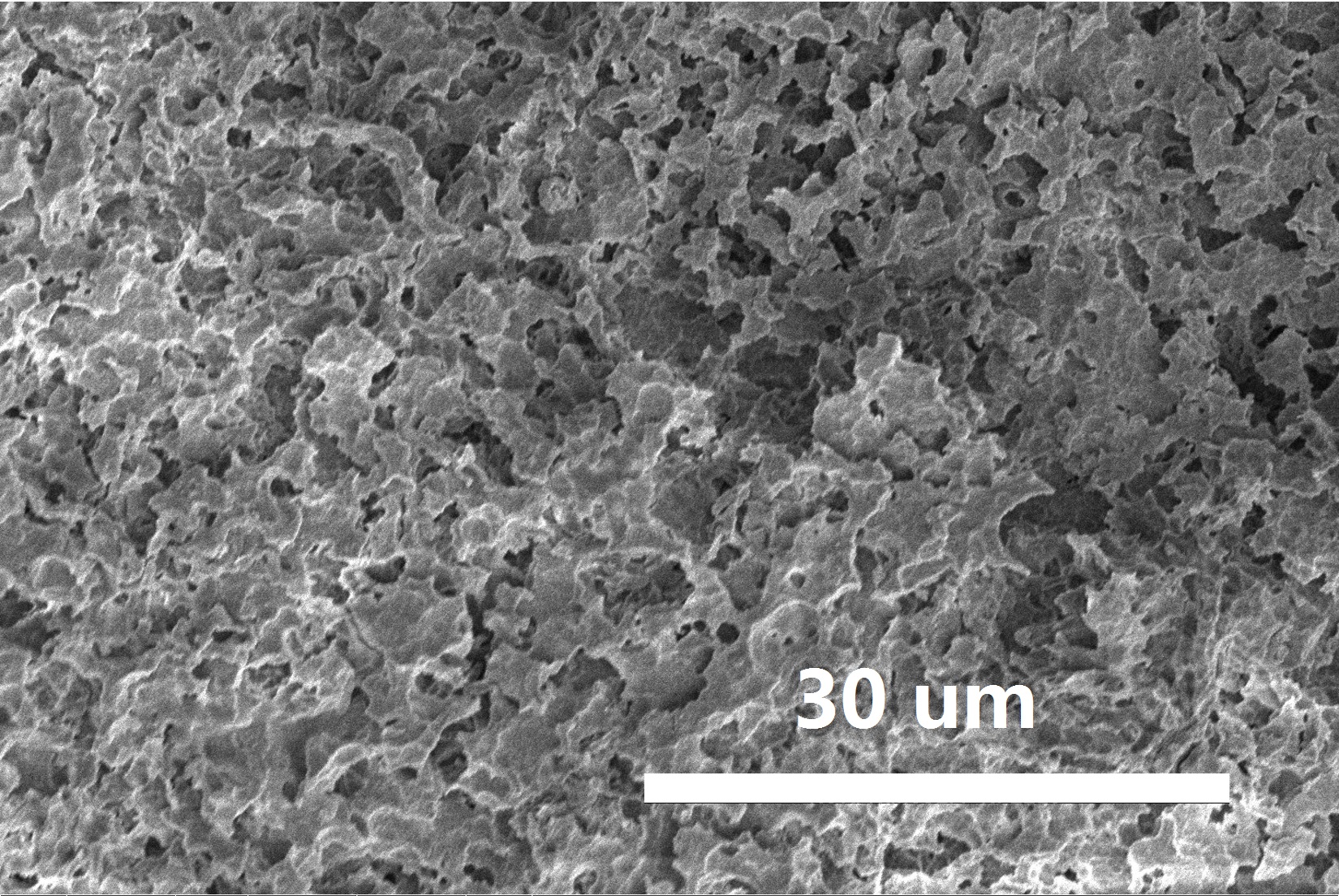}
	}\\
	\subfloat[]{
	\includegraphics[width=0.48\textwidth]{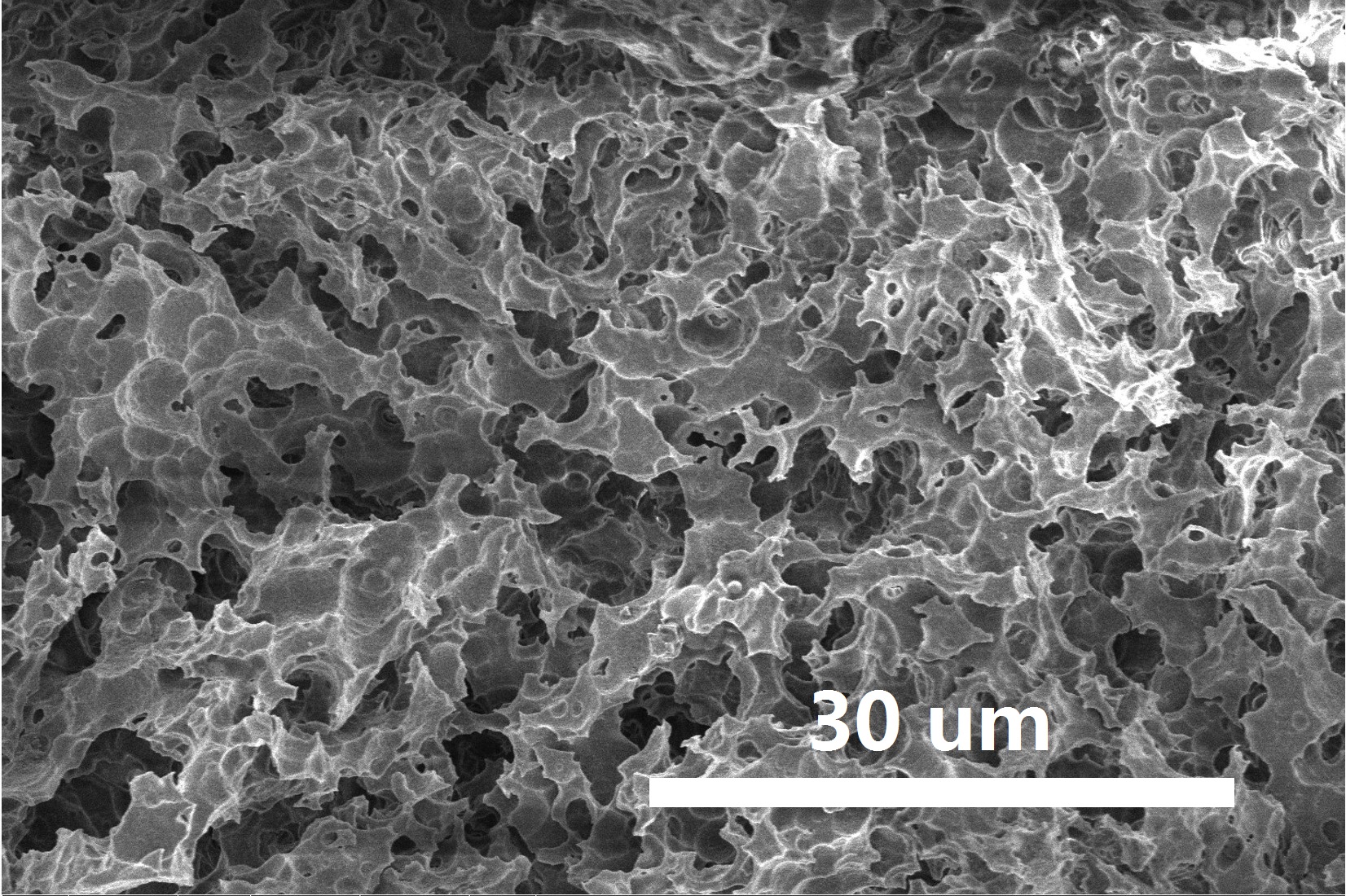}
	}\\
   }
    \caption{The SEM picture of quenched PLA/PEG blend after $160{h}$ degradation of (a) 90/10 and (b) 70/30.}
    \label{fig:SEMque}
\end{figure}

Figure \ref{fig:SEMque} is the SEM of quenched samples B' and D'. We can clearly observe a porous foam structure. Because the heating process induced the phase separation. and the PEG portions not entangled with PLA matrix could easily solve out, and left the quenched PLA. This observation agrees with the initial sharp weight loss in the degradation behavior in Fig.3b. And the porous structure also provided larger surface area for a faster degradation rate following the initial weight loss. The sample D' as shown in Fig.\ref{fig:SEMque}b has similar porous feature, except the structure is more fragmentary with more PEG solved out. 

\begin{figure}
    \centering{
	\subfloat[]{
		\includegraphics[width=0.48\textwidth]{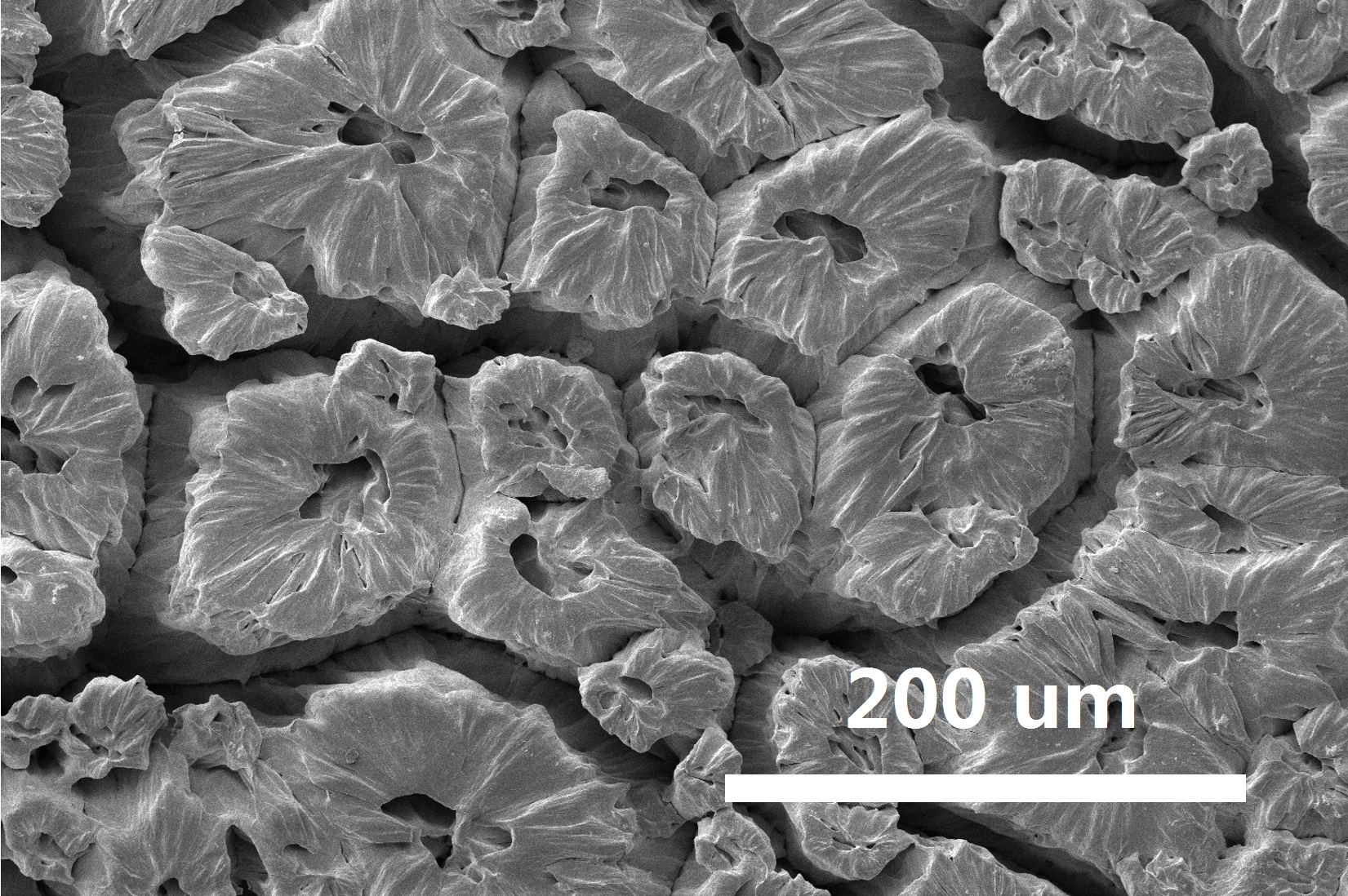}
		}\\
	\subfloat[]{
		\includegraphics[width=0.48\textwidth]{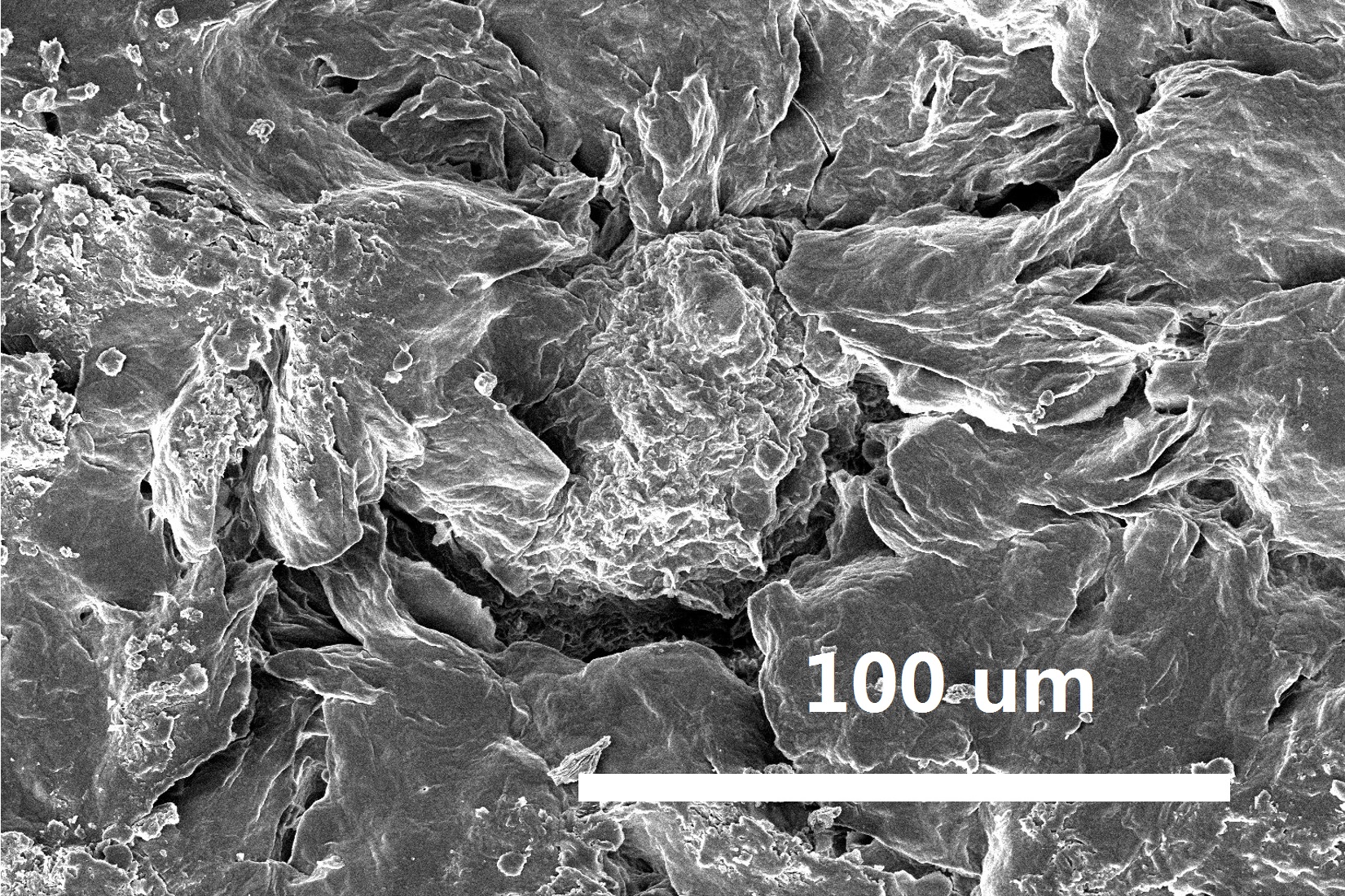}
		}\\
	\subfloat[]{
		\includegraphics[width=0.48\textwidth]{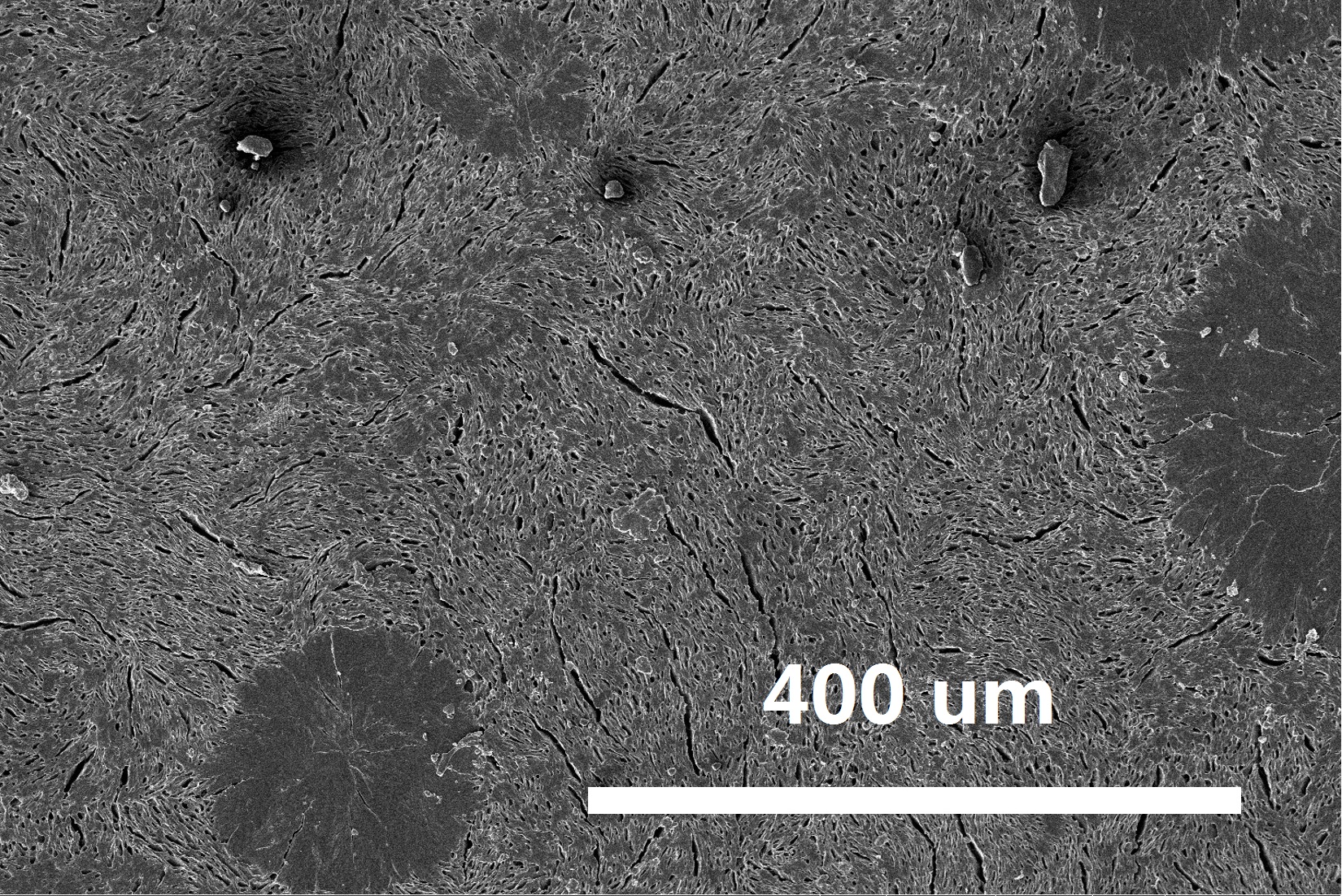}
		}\\
    }
    \caption{The SEM picture of recrystallized PLA/PEG blend of (a) 90/10 and (b) 80/20 after $160{h}$ degradation, and (c) 70/30 after $6{h}$ immersion in NaOH solution. }
    \label{fig:SEMcrys}
\end{figure}

Figure \ref{fig:SEMcrys}a is the SEM of sample B after $160h$ immersion in base solution. The intriguing `islands' morphology indicates the phase separation during reheating, and the PLA self-aggregation to form spherulites during slow cooling. The slits between PLA `islands' are not caused by either degradation or the thermal contraction, instead these gaps were originally occupied by the PEG portions and caved out by the PEG solving out. 

Unlike in sample B, where the PLA portion trends to aggregate together and squeeze the PEG to the surface to line out the `islands', in sample C the relatively larger PEG content can penetrate into the matrix and form deeper gaps. Figure \ref{fig:SEMcrys}b shows such deeper cracks and a PLA spherulite is lined out by the cracks. It implies that if the PEG wt\% is higher, this PLA portion can be completely separated out from the matrix, and disperses into the environment with PEG solving. 

To understand what happened during the striking dispersion, sample D were observed after $6{h}$ immersion as shown in Fig.\ref{fig:SEMcrys}c. Except several PLA spherulites on the surface, other area is fully marked with small slits and cracks indicating the PEG is solving out. Since the PEG wt\% are as high as to separate those PLA spherulite crystals, the solving of PEG spherulite implies that the total dispersion is approaching. (This sample dispersed in $\sim7.5{h}$ in the following degradation experiment.)

For a short summary, the SEM results confirm that: 1) The heating process induces a phase separation between PLA and PEG, and sufficient recrystallization time allows the polymer chain to self-aggregate, leading to a subsequent larger phase separation; 2) Phase separation is the reason behind the `striking dispersion' we have observed. The PLA does not really degrade in such a short time, instead the very small PLA fragments are dispersed into the solution after neighboring PEG portions solved out. This phase separation and PEG solving effect actually exists in all samples with various PEG wt\%, evidenced by the initial sharp weight loss in the degradation experiments (Fig.3). And when the PEG content is as high as 30\%, this effect breaks and disperses the entire matrix.
   
\subsection{Effect of Miscibility}

Miscibility effect is a fair guess on the reason that this fast striking dispersion/degradation has not been reported in others' studies on the PEG-PLA mixture. Previous PEG modifications on PLA usually employed the melt twin screw extruding technique, which leads to a better mixture, and the mixing is done at high temperature, i.e. the chains already experienced a rearrangement process while being mixed. The miscibility issue of PLA and PEG blend has been well studied by others and many reports concluded that the miscibility is not good and analyzed the reason from different aspects. Our work of the aqueous degradation provides an alternative evidence that the wettability could be an considerable reason of the poor miscibility.

Another possible reason that this phenomenon has not been reported, is that the studies on recrystallization and phase-separation did not extend their researches on the degradation, and the studies on degradation did not involve a recrystallization treatment, Ref. \cite{Serra,Sheth} as examples. However in these two papers we can still infer the clues of dispersion. In Ref. \cite{Serra} the in-vitro degradation studies showed that the inclusion of PEG significantly accelerated the degradation rate of the PLA-based 3D printed tissue engineering scaffolds. This work employed PEG/PLA solution blend, and there is no thermal process so we assume the materials is amorphous. The authors reported that a striking degradation was observed in the case of the blend with 20\% PEG after two weeks of immersion in Simulated Body Fluid. The hydrophilic PEG domains in the films surface led to a faster degradation when it is in contact with the fluid. In our opinion, this is not an actual degradation and it is caused by the PEG being dissolved.

In Sheth et al.'s paper \cite{Sheth}, their results displayed an abnormally high degradation rate of high PEG sample (70\%), which could be very likely the same phenomenon we observed. The author correctly concluded the PEG dissolution but did not realize the PLA dispersion along with the dissolution of PEG. They also have found the miscibility of PLA/PEG blends varies the degradation rate, which agrees with our results.

\section{Conclusion}

A striking dispersion behavior of recrystallized PEG/PLA solvent-casting blend of 30/70 weight ratio was reported in this work. The materials has been observed to entirely dissolve in 0.1 mol/L NaOH solution in around $6$-$8{h}$. The reason of this phenomenon was hypothesized to be the phase separation and PEG dissolving. The recrystallization process forms large PLA self-aggregation, separated by the PEG portions. When immersed in aqueous environment the PEG solved out in around $6{h}$ and the PLA portions were subsequently dispersed into the solution, then the entire matrix broke and it look like a total degradation. Following the same underlying reason, the degradation behaviors of the initial sharp weight loss and the step weight loss on low PEG wt\% samples can also be understood in this way.  

Characterizations have been done to validate the hypothesis. The XRD results indicate a semi-crystal PLA and amorphous PEG mixture for the recrystallized blend. The DSC results support the PLA crystal confinement to PEG portions, and evidents the phase separation. The SEM provides a direct observation on the morphology of each sample and proves the hypothesis. In this way, we have concluded that the poor miscibility of PEG and PLA is an important factor of the thermal-induced phase separation and the subsequent striking dispersion.


\begin{thebibliography}{43}

\bibitem {rev2} Nampoothiri, K. M.; Nair, N. R.; John, R. P. \textit{Bioresour. Technol.} 2010, \textbf{101}, 8493. 

\bibitem {rev4} Lim, L. T.; Auras, R.; Rubino, M. \textit{Prog. Polym. Sci.} 2008, \textbf{33}, 820. 

\bibitem {rev5} Bordes, P.; Pollet, E.; Av\'{e}rous, L. \textit{Prog. Polym. Sci.} 2009, \textbf{34}, 125. 

\bibitem {rev6} Rasal, R. M.; Janorkar, A. V.; Hirt D. E. \textit{Prog. Polym. Sci.} 2010, \textbf{35}, 338. 

\bibitem {pk1} Yuk, K. Y.; Choi, Y. M.; Park, J. S.; Kim, S. Y.; Park, K. N.; Huh, K. M. \textit{Polymer(Korea)} 2009, \textbf{33}, 469. 

\bibitem {Sheth} Sheth, M.; Kumar, R. A.; Dave, V.; Gross, R. A.; McCarthy, S. P. \textit{J. Appl. Polym. Sci.} 1997, \textbf{66}, 1495.
 
\bibitem {Serra}  Serra, T.; Ortiz-Hernandez, M.; Engel, E.; Planell, J. A.; Navarro, M. \textit{Mater. Sci. Eng. C} 2014, \textbf{38}, 55.

\bibitem{13} Hernandez-Montero, N.; Ugartemendia, J. M.; Amestoy, H.; Sarasua, J. R. \textit{J. Polym. Sci. Polym. Phys.} 2014, \textbf{52}, 111.

\bibitem{17} Pillin, L.; Montrelay, N.; Grohens, Y. \textit{Polymer} 2006, \textbf{47}, 4676. 

\bibitem{24} Pereira, A. G.; Gouveia, R. F.; de Carvalho, G. M.; Rubira, A. F.; Muniz, E. C. \textit{Mater. Sci. Eng. C} 2009, \textbf{29}, 499. 

\bibitem{25} Zhang, Y.; Wang, Z.; Jiang, F.; Bai, J.; Wang, Z. \textit{Soft Matter} 2013, \textbf{9}, 5771. 

\bibitem{36} Averous, O. M. \textit{Polymer} 2001, \textbf{42}, 6209. 

\bibitem{33} Qiu, Z.; Ikehara, T.; Nishi, T. \textit{Polymer} 2003, \textbf{44}, 3101. 

\bibitem{pk2} Yoon, C. S.; Ji, D. S. \textit{Polymer(Korea)} 2009, \textbf{33}, 581. 

\bibitem{Albertsson1} K\"{a}llrot, M.; Edlund, U.; Albertsson, A. \textit{Biomacromolecules} 2007, \textbf{8}, 2492. 

\bibitem{deg2} Tsuji, H.; Ikada, H. J. \textit{J. Appl. Polym. Sci.} 1997, \textbf{63}, 855.
\bibitem{cooper} Heath, D.; Cooper, S. \textit{J. Biomed. Mater. Res. A} 2010, \textbf{94}, 1294.

\bibitem{RanML} Huang, R.; Zhu, X.; Tu, H.; Wan, A. \textit{Mater. Lett.} 2014, \textbf{136}, 126.

\bibitem{RanMRX} Huang, R.; Zhu, X.; Zhao, T.; Wan, A. \textit{Mater. Res. Express} 2014, \textbf{1}, 045403.

\bibitem{pegXRD} Wang C.; Feng, L.; Yang, H.; Xin, G.; Li, W.; Zheng, J.; Tian, J.; Li, X. \textit{Phys. Chem. Chem. Phys.} 2012, \textbf{14}, 13233. 

\end{thebibliography}
\end{document}